\documentstyle[12pt]{article}
\begin{document}
\setlength{\baselineskip}{2em}
\begin{titlepage}
\begin{center}
{\Large\bf Bose-Einstein Condensation in a Constant 
Magnetic Field } \vskip 1.0 cm

{\bf H. Perez Rojas}$^{1,2,3}$,
{\bf L. Villegas-Lelovski}$^{2}$.
\vskip 1.5cm
$^1${\it High Energy Division, Department of Physics, University of
Helsinki ,\\
Siltavuorenpenger 20C, FIN-00014, University of Helsinki, Finland}
\\
$^2${\it Departamento de F\'{\i}sica, Centro de Investigaci\'on y de
Estudios Avanzados del IPN,\\
Apartado Postal 14-740, 07000 M\'exico, D. F., M\'exico}\\
$^3${\it Grupo 
de Fisica Te\'orica ICIMAF,}\\
{\it Calle E No. 309,Vedado, La Habana 4, Cuba.}
\end{center}


\begin{abstract}                                                             
\noindent
We discuss the
occurrence of  Bose-Einstein 
condensation in  systems of noninteracting charged particles in three 
in one dimensions and in presence
of an external magnetic 
field. In the one dimensional, as well as
in the magnetic field cases, 
although not a critical temperature, a characteristic temperature can be 
found, corresponding to the case in which the ground state density becomes 
a macroscopic fraction of the total density.
The case of relativistic charged scalar and vector particles is
studied. The results obtained
give support to the existence of superconductivity in extremely strong
magnetic
fields, and leads to  the prediction of
superconductive-ferromagnetic behavior in the vector
field 
case, which might be of interest in condensed matter as well
as in cosmology. Some features of the magnetization in the early universe 
are conjectured.
\end{abstract} 
\noindent
\end{titlepage}

\newpage
\section{Introduction}
Condensation in the ground state seems to be a general property whenever
the conditions of quantum degeneracy of the Bose-Einstein gas are
satisfied. 
Quantum degeneracy is usually understood to be achieved when the De
Broglie thermal wavelength $\lambda$ is greater that the mean
interparticle
separation $N^{-1/3}$. The remarkable discovery made by Einstein on
the Bose distribution was that condensation may occur starting at some
critical temperature $T_c$ different from zero, which is usually
referred as Bose-Einstein condensation (BEC). Condensation of charged
particles requires always from a background of charge of opposite sign
to screen the mutual repulsion among particles. We will assume that
such background exists.

The condition for  BEC to occur in a gas of charged particles placed in a
magnetic field
is usually borrowed from the case without magnetic field as $\mu = E_0$
(where $E_0$ is the ground state energy), which can be translated into
$\mu = \sqrt{ M c^2 \pm e B \hbar 
c}$ 
for the relativistic case ($\mu' = \mu - M c^2 \pm e B \hbar/M c$ in the 
nonrelativistic limit). 
We must cite first Schafroth  \cite{Schafroth} who proved
that for a 
non-relativistic boson gas, Bose-Einstein condensation (BEC)
does not occur in presence of a constant magnetic field. It was implicit
in his paper that he considered only the case of not very intense fields. 
The problem was studied afterwards by May \cite{May}, who 
showed that a phase transition like that in the free gas occurs for 
$D \geq 5$, and later by 
Daicic {\it et al} \cite{Daicic}, 
\cite{Gailis}, who extended the considerations made by \cite{May} to the 
relativistic high temperature case. 
Later Toms \cite{Toms} proved 
that  BEC in presence of a constant magnetic field does not occur in any 
number of spatial dimensions, and Elmfors {\it et al} \cite{Elm} 
who stated that in the 3$D$
case, although a true condensate is not formed, 
the Landau ground state can be occupied by a large charge density.
A simple criterion for BEC to occur, was given by Toms and Kirsten 
\cite{Toms}, who concluded that usual BEC can occur only for $d \geq 3$.

There are two ways of 
characterizing the occurrence of BEC,
which are: 
1) The existence of a critical temperature $T_c > 0$ such
that $\mu(T_c) = E_0$, (This condition is usually taken as a necessary and 
sufficient
condition for condensation; see  \cite{Kirsten}). Then for
 $T \leq T_c$, some significant amount of particles starts to condense 
in the ground
state. 2) The existence of a finite fraction of the total particle density
in
the ground state and in states in its neighborhood at some
temperature $T > 0$. We refer to 1) as the strong and to 2) as the weak
criterion for BEC.

A magnetic field breaks explicitly the spatial symmetry.  It is reflected
in the wave function of a
charged particle
moving 
in it, and also in its spectrum. The spectrum of a charged Fermi or Bose 
particle
in a constant magnetic field indicates this breaking of the symmetry in 
momentum space: the momentum components perpendicular to the field, 
collapse in a set of discrete Landau quantum states, their energy
eigenvalues having infinite degeneracy.  A gas of charged particles
(either Bosons or Fermions) in presence of very intense
magnetic fields populate mainly the Landau ground state
$n = 0$ 
and behaves as a one-dimensional gas in the axis $p_3$ parallel to the 
field. For Bosons, it leads to a statistical distribution  for fixed 
temperature $T$ and magnetic field $B$, which depends only
on $p_3$.

We must 
bear in mind at this point that the usual properties of second order phase
transitions
cannot be valid in presence of  external 
fields (see Landau-Lifshitz, \cite{Landau}). The external field introduces 
in the Hamiltonian a perturbing operator
which is linear in the external field strength, in our case $B$, and in
the order operator , in our case
$\hat {\cal M}$, the magnetization. As a result, at any value of
$B$, ${\cal M}$
becomes different from zero at any temperature. Thus, $B$ reduces the
symmetry 
of the usual more symmetrical phase, and the difference between the two
phases 
of the usual theory dissapears; the discrete transition point dissapears;
the 
transition is "smoothed out".

Thus, as BEC in 3D has the properties of a second order phase transition,
in
studying it in presence of a magnetic 
field, the strong criterium cannot be applied; no critical temperature 
separating phases of different symmetry
exists, and that was confirmed in many of the results of references
\cite{Schafroth}-
\cite{Toms}. But an order parameter exists at any temperature, the 
magnetization ${\cal M}$, which for large fields and densities, and
temperatures small 
enough (see below) becomes proportional to the ground state density.
Actually,
the ground state density can be considered also as another order parameter 
closely related to ${\cal M}$ and 
may become
a macroscopic fraction of the total density, not at a definite $T_c$, but
in 
some interval of values of the temperature. 
We may think in the case in which one starts from a gas of charged
particles, without
magnetic field, below the critical
temperature. If a sufficiently strong magnetic field is applied, the
particles in the condensate remain condensed, and a critical temperature
is not possible to be defined.
(It is interesting to mention here that, as shown in ref. \cite{Toms1},
there 
is no critical temperature also in the case of a gas confined to a 
parabolic potential.)

This paper 
is a more detailed and complete version 
of some of the results of two  previous ones, \cite{HPerez}, \cite{Hugo}, 
in which it was
pointed out that 
Bose-Einstein condensation, in the sense of
having a large 
population in the ground state in a continuous 
range of temperatures (i.e., no critical temperature exists), 
may occur in presence of a magnetic field {\bf B}, 
for very dense systems at very low temperatures.
The existence of discrete Landau quantum states allows this
"weak" BEC. In this case, the ground state population is explicitly
included in the Bose-Einstein distribution. 

The occurrence of Bose-Einstein condensation in a strong magnetic
field gives theoretical support to the phenomenon of re-appearance of
superconductivity in fields
strong enough, and even leads to  the prediction of
superconductive-ferromagnetic behavior in the vector
field case, a phenomenon which would be of especial interest in condensed
matter
and in cosmology.

The structure of the present paper is as follows:
in section 2 we shall discuss some features of the usual $3D$
Bose-Einstein condensation, which gives a basis to understand its
"weak" occurrence in the one-dimensional case (section 3) and in the
magnetic
field
case (section 4). In section 5 it
is explicitly shown the ferromagnetic behavior of the magnetization of the
vector field case. It is also discussed the connection with
superconductivity, the existence of low-field and
strong-field condensation and the corresponding superconductive behavior,
separated by a gap,
and the ferromagnetic-superconductive behavior appearing in the vector
field case.
Sect 6 is devoted to some cosmological considerations, and section 7 deals
with
the conclusions.

\section{The usual Bose-Einstein Condensation}

In order to be self-contained, we recall some results and formulae
from \cite{HPerez}, \cite{Hugo}.
concerning the standard $3D$ BEC theory of the free Bose gas. We must
emphasize that our
considerations in the present and next section are essentially
concerned
with the case of systems far from the thermodynamic limit. We
shall use throughout the paper $T$ in energy units, i.e., as the
product of the Boltzmann constant $k$ by the absolute temperature; thus,
$T
= \beta^{-1}$. The
chemical potential $\mu =f(N,T)<0$ is a
decreasing function of temperature at fixed density $N$, and for $\mu =0$
one gets an equation defining $T_c = f_c(N)$. For temperatures $T < T_c$,
as $
\mu = 0$, the expression for the density gives values $N^{\prime }(T) <
N$, and
the difference $N - N^{\prime} = N_0$ is interpreted as the density of
particles in the condensate. The mean interparticle separation is then 
$l=N^{-1/3}$.

In our considerations we will use integrals which, as it is usually done, 
must be interpreted as approximations of sums over discrete quantum
states, 
which imply not to be working at the thermodynamic limit. For usual 
macroscopic systems, as the separation between quantum states is 
$\Delta p = h/V^{1/3}$,
the approximation of the sum by the integral is quite well justified.

Now, above the critical temperature for condensation 
\begin{equation}
N=4\pi^{-1/2} \lambda ^{-3}\int_0^\infty \frac{x^2d}{e^{x^2 + \bar \mu
}-1} = \lambda ^{-3}g_{3/2}(z),
\end{equation}
\noindent
where $\bar \mu =-\mu /T(>0)$, $x=p/p_T$ is the relative momentum, $p_T=
\sqrt{2mT}$ the characteristic thermal momentum, and $\lambda =h/(2\pi
mT)^{1/2}$ the de Broglie thermal wavelength. The function $g_n(z)$ is 
easily defined (see Pathria \cite{Pathria}), 
as a function of  the fugacity $z=e^{\mu /T}$,. At $T=T_c$, we have
$g_n(1)=\zeta (n)$, 
and $g_{3/2}(1)=\zeta (3/2)$, and the density is $N_c=\zeta
(3/2)/\lambda^3$,
or in other words, $N\lambda^3=\zeta (3/2)\simeq 2.612$. 
Here we must point out that if the particle density is written as a sum
over quantum states 
\begin{equation}
N = \sum_i (z^{- 1} e^{\frac{E_i}{T}} - 1)^{- 1}, \label{Path}
\end{equation}
for
strictly $z =1$ ($\mu = 0$), the ground state term diverges. (In field
theoretical language, it is due to an infrared divergence of the
one-particle Green
function at $\mu = 0$). This divergence
usually serves as an argument \cite{Pathria} to indicate the occurrence of
Bose-Einstein condensation, and actually we must take $\mu = 0_+$.
However, in considering temperatures below $T_c$, 
it is a very good approximation to take $\mu = 0$ \cite{Pathria},
\cite{Landau}, 
and thus the
ground state density is described by the expression 
\begin{equation}
N_0 = N [1 -(\frac{T}{T_c})^{3/2}] \label{Ter}
\end{equation}
It leads to conclude that for $T =
T_c$, $N_0 = 0$. This seems to be
rather inconsistent with our previous analysis about the limit of
(\ref{Path}) for $\mu
\to 0$,
and one expect that (\ref{Ter}) is not exact, but must contain terms
of order smaller than $N$, accounting for the ground state density at
$T = T_c$. 
One can proceed in a different way, by keeping in mind the temperature 
Green's function formalism \cite{Fradkin}, which in the non-relativistic
infinite
volume limit leads
to $N = \int d^{3}{\bf p} G(p)$. By considering a finite
volume, $N$ results as a sum over
degenerate energy states, the degeneracy factor being proportional to $p^2
\Delta p/h^3$. This leads to a {\it finite} contribution in the limit $p =
0$, since the infrared divergence of the Green function at the pole 
$\mu - E_0$ is canceled by the factor $p^{2}$, leading to a non-zero
but {\it finite}
density at the ground state at $T = T_c$, as will be seen below.

Before doing that, it is interesting to investigate in detail the particle
density in relative
momentum space 
\[
f_3(x,\bar \mu )=\frac{x^2}{e^{x^2+\bar \mu }-1}. 
\]
By calculating the first and second derivatives of this function, we find
that for $\bar \mu \neq 0$  it has a minimum at $x=0$ and a maximum at $
x=x_\mu $ where $x_\mu $ is the solution of $e^{x^2+\bar \mu }=
1/(1-x^2)$.

In this sense $f(x,\bar \mu )$ has a similar behavior than the
Maxwell-Boltzmann distribution of classical statistics. But as $\bar \mu
\to
0$, also $x_\mu \to 0$, and the maximum of the density, for strictly $\bar
\mu =0$, is located at $x=0$. The convergence to the limit $x=0$ is not
uniform. A finite fraction of the total density falls in a vicinity of the
ground state.
This behavior is illustrated in Fig. 1.

If we go back and substitute the original integral over momentum by a sum
over shells of quantum states of momentum (energy states), we can write
for the critical value $z = 0$,
\begin{equation}
N_c = \frac{4 \pi}{h^3} \sum_{i = 0}^{\infty} \frac{p^2_i \Delta p}{
e^{p_i^2/2 M T} - 1}.
\end{equation}
\noindent
Taking $\Delta p \simeq h/V^{1/3}$, where $V$ is the volume
of the vessel containing the gas, the contribution of the first term
$p_0 = 0$ of the ground state
density is $N_0 = \frac{4}{V^{1/3} \lambda^2}$. We have thus a fraction of 
\begin{equation}
N_0/N_c = 4 \lambda \zeta (3/2)/V^{1/3} = 4 \zeta (3/2)^{-2/3}/{\cal
N}^{1/3}
\label{sen}
\end{equation}
\noindent
particles in the ground state, (where ${\cal N} = N V$ is the total
number of particles) which is the most populated, as described by
the statistical distribution, at $T = T_c$. A numerical estimation for one
litre of $He$ gas leads to $N_0/N \simeq 10^{-6}$. In quantum states in a
small neighborhood of the ground state, the momentum density has slightly
lower values. Thus, at the critical temperature for BEC, there is a set of
states close to the ground state, having relative large densities.
Although this result was obtained using exactly $\bar\mu = 0$, which
according to (\ref{Path}) implies an infinite ground state population, it
represents an improvement compared with $N_0 = 0$ for $T = T_c$.

Eq. (\ref{sen}) indicates an interesting relation: the larger the
separation
between quantum states, (i.e., the smaller the volume) the larger the
population of the ground state at the
critical temperature. In the thermodynamic limit the quantum states form a
continuum, and (\ref{sen}) has no meaning. However, most systems of
physical
interest, in laboratory as well as in astrophysical
contexts, have finite $V$ and ${\cal N}$.

We conclude that for values of $T<T_c$, the curve describing the density
in momentum space
flattens on the $p$ (or $x$) axis, and its maximum decreases also. As 
conservation of particles is assumed, we get the ground state
density by adding to $f(x)$ the quantity $2N[1-(T/T_c)^{3/2}]\theta
(T_c-T)\lambda ^3\delta (x)$ as an additional density. As $T\rightarrow
0$,
the density in the ground state increases at the expense of non-zero
momentum states. This leads to the usual Bose-Einstein condensation. We
must
remark that in that case, for values of $T$ smaller but close to $T_c$,
both
the weak and the strong criteria are satisfied. 

\section{The one-dimensional case}

The condensation in 1D is interesting in itself, since there is an
incresing
 interest to study condensation in systems of lower dimensionality than 3D 
(we discussed the ideal 2D gas case in \cite{Hugo}), taking into account
its possible 
experimental realization (even the problem 
of superfluidity in quasy 1D systems is currently discussed in the
literature, 
see \cite{Kagan}). Here we will be interested on it thinking in its
connection 
with the magnetic field case.
For $D = 1$, there is reduced
symmetry with regard to the 3D case, and no critical temperature is
expected 
to occur. In this case, the mean interparticle
separation is $l = N^{-1}$. The density in momentum space coincides with
the
Bose-Einstein distribution, $f_1 (p, T, \mu) = (e^{p^2/2M T - \mu/T} -
1)^{- 1}$.
This function has only one extremum, a maximum, at $p = 0$.  We have the 
expression for the density of particles as

\begin{equation}
N = 2 \lambda^{-1}\pi^{- 1/2} \int_0^\infty \frac{d x}{e^{x^2 + \bar \mu}
- 1
}= \frac{1}{\lambda} g_{1/2} (z).
\end{equation}

We have thus that $N\lambda = g_{1/2}(z)$. The fact that $g_{1/2}(z)$ 
diverges as $z \to 1$ indicates an enhancement of the quantum
degeneracy regime. But this fact actually means that $\mu $ is a
decreasing
function of $T$ for $N$ constant, and for very small $\bar \mu $ one can
write, approximately 
\begin{equation}
N\simeq 2\lambda ^{-1}\pi ^{-1/2}\int_0^{x_0}\frac{dx}{x^2+\bar \mu
}\simeq 
\frac{\pi ^{1/2}}{\bar \mu ^{1/2}\lambda },  \label{1}
\end{equation}
where $x_0=p_0/MT$, $p_0$ being some characteristic momentum $p_0\gg p_T$.
Thus, $\bar \mu $ does not vanish at $T\neq 0$, and for small $T$ it is
approximately given by $\bar \mu =\pi /N^2\lambda ^2$. 
Then $\bar \mu \ll x_0$ implies $p_0 \gg 2 \pi^2 M^2 T^2/N^2 h^3$, and it
is fulfilled if $T/N \to 0$.

By substituting the last expression for $\bar \mu $ back in (\ref{1}), one
has 
\begin{equation}
N \simeq 2\lambda ^{-1}\pi^{-1/2}\int_{-x_0}^{x_0}\frac{dx}
{x^2+\gamma ^2},
\end{equation}
where $\gamma =\pi ^{1/2}/N\lambda $. Due to the properties of the Cauchy
distribution, one can find that one half of the density is in the interval 
$[0, \gamma]$,
\[
\frac{1}{2}N = 2\lambda^{-1} \pi^{-1/2} \int_0^\gamma 
\frac{ dx}{x^2+\gamma ^2}.
\]
We can also find
\[
f_1(p,T,N)_{\gamma \to 0}\simeq \frac{2N \lambda}{\pi^{1/2}} \delta (p).
\]
Thus, for small $\gamma$ one-half of the density is concentrated in the 
interval of momentum $[-\gamma, \gamma]$. To precise figures, we shall
define $\gamma' =\gamma p_T$,
where $p_T=\sqrt{2\pi M T}$
is the de Broglie momentum, to give the proper 
dimensions, and compare it with the ground state quantum cell, 
$2\pi\hbar/L$, where $L$ is the one dimensional volume of our system. For 
$2\pi\hbar/L \gg \gamma'$, most of the system is in the ground state. 
This means that the adimensional phase space density is much greater 
than the "volume" of the system measured in $\lambda$ units, 
$N\lambda \gg  L/\lambda$. 

For $2\pi\hbar/L \simeq \gamma'$, $N\lambda \sim L/\lambda$, and we can
define a temperature $T_d= N
h^2/2 \pi^{3/2} mL$ which, although not being a critical
temperature for a
phase transition, establishes the order of magnitude of $T$ in which the 
ground state density is a macroscopic fraction of the total density $N$. 
Thus, in the present case we have a weak condensation.

\section{The magnetic field case}

By assuming as in \cite{Elm} a constant microscopic magnetic field $B$ 
along the $p_3$ axis
(the external field is $H^{ext} = B - 4 \pi {\cal M}(B)$, where ${\cal 
M}(B)$ is the magnetization), for $eB \hbar/m c \gg T$,  all the density
can be taken as concentrated in the Landau state $ n = 0$, the maximum of
the
density in momentum space is just at the point $x = 0$ {\it even for $\mu' 
\neq 0$}; thus, we conclude that without any critical temperature 
in that case
a finite fraction of the density is found in the ground state. 

The energy of a charged particle in a magnetic field in the 
non-relativistic case is $p_3^2 / 2 M + e B \hbar(n + \frac{1}{2})/M c$.
Here
we name $\mu_1 = e B \hbar/2 M c - \mu$ as the effective chemical
potential.
By defining the elementary cell as $v = \lambda h c/e B$ we can write
\begin{eqnarray}
N &=& \frac{1}{v} \int_0^{\infty} \frac{d x}{e^{x^2 + \bar \mu_1} - 1}
\\[1em]
&&\mbox{} \frac{1}{v} g_{1/2} (z) =\frac{1}{v} \sum_{m =
1}^{\infty}\frac{e^{-\mu_1 m \beta}}{m^{1/2}}
\end{eqnarray}

By introducing the continuous variable $x = T m$ and by writing $T 
\sum_{n=1}^{\infty} \to \int_0^{\infty}$ we get
\begin{equation}
N = \frac{\pi^{1/2}}{ v \bar \mu_1^{1/2}} ,
\end{equation}
\noindent
and in a similar way as before one can get for the relative population in 
the ground state
\begin{equation}
\frac{N_0}{N} = \frac{\lambda}{ L \bar \mu_1^{1/2}}. \label{l}
\end{equation}
Here $L$ is the characteristic dimension of the system (along the magnetic 
field). We observe that $N_0/N$ increases with decreasing $\bar \mu_1$,
i.e.,
as temperature decreases. The limit, which is not transparent from
(\ref{l}) 
must be unity, as we shall see below. We conclude that the density in 
momentum space in the magnetic field case concentrates in a (decreasing 
with T) narrow peak around the ground state $p_3 = 0$ (Fig. 2).

We turn  to the relativistic case.
The thermodynamic potential for a gas of charged scalar particles placed
in 
the magnetic field is ,
\begin{equation}
\Omega_s = \frac{eB}{4 \pi^2 \hbar^2 c \beta} \sum_{n = 0}^{\infty} 
\int_{-\infty}^{\infty} dp_3 \left[\ln 
(1 - e^{-(\epsilon_q - \mu)\beta})(1 - e^{-(\epsilon_q + \mu)\beta}) 
+ \beta \epsilon_q \right] ~\label{O}
\end{equation}
\noindent
where $\epsilon_q = \sqrt{ p_3^2 c^2 + M^2 c^4 + 2 eB \hbar c (n + 
\frac{1}{2})}$, the last term in (1) accounts for the vacuum energy
and $\mu$ is the chemical potential. 
For a vector field the one-loop thermodynamic potential is
\begin{eqnarray}
\Omega_v& =& \frac{eB}{4 \pi^2 \hbar^2 c \beta} 
\int_{-\infty}^{\infty} dp_3 \left[\ln 
(1 - e^{-(\epsilon_{0q} - \mu)\beta})(1 - e^{-(\epsilon_{0q} + 
\mu)\beta}) + \beta \epsilon_q \right] \nonumber
\\[1em]
&&\mbox{}
+ \frac{eB}{4 \pi^2 \hbar^2 c \beta} \sum_{n = 0}^{\infty} 
\beta_n \int_{-\infty}^{\infty} dp_3 \left[\ln 
(1 - e^{-(\epsilon_q - \mu)\beta})(1 - e^{-(\epsilon_q + \mu)\beta}) 
+ \beta \epsilon_q \right]
\end{eqnarray}
\noindent
where $\beta_n = 3 - \delta_{0n}$, $\epsilon_{0q} = 
\sqrt{ p_3^2 c^2 + M^2 c^4 -  eB \hbar c}$, $\epsilon_q = 
\sqrt{ p_3^2 c^2 + M^2 c^4 + 2 eB \hbar c (n + \frac{1}{2})}$.

The mean density of particles minus antiparticles (average charge 
divided by $e$) is given by $ N_{s,v} = - \partial \Omega_{s, v}/ 
\partial \mu $.

We have explicitly
\begin{equation}
N_s = \frac{e B}{4 \pi^2 \hbar^2
c}\sum_0^{\infty}\left[\int^{\infty}_{-\infty} 
dp_3 (n_p^+ - n_p^-) \right].
\end{equation}
\noindent
where $n_p^{\pm} = [exp(\epsilon_q \mp \mu)\beta - 1]^{-1}$.

We take the expression fro the vector field case from the Weinberg- Salam
Lagrangian in a medium in a magnetic field \cite{Perez}. We have, 
\begin{eqnarray}
N_v & =&\frac{e B}{4 \pi^2 \hbar^2 c}\left[\int^{\infty}_{-\infty} 
dp_3 (n_{0p}^+ - n_{0p}^-) \right]\nonumber
\\[1em]
&&\mbox{} + \frac{e B}{4 
\pi^2 \hbar^2 c}\sum_0^{\infty}\beta_n \left[\int^{\infty}_{-\infty} 
dp_3 (n_p^+ - n_p^-) \right]
\end{eqnarray}
\noindent
where $n_{0p}^{\pm} = [exp(\epsilon_{0q} \mp \mu)\beta - 1]^{-1}$, 
$\epsilon_{0q} = \sqrt{ p_3^2 c^2 + M^2 c^4 - e B \hbar c}$, and 
$\beta_{0n} = 3 - \delta_{0n}$. 

For $T \to 0$, $\mu \to M_{\pm} c^2$  (we named 
$M_{\pm} = \sqrt{M^2 \pm e B \hbar/c^3}$), 
the population in Landau quantum states other than $n = 0$ vanishes (this 
was shown in [1]) and the density  for the $n = 0$ 
state  is infrared divergent. We expect then most of the population to 
be in the ground state, since for small temperatures $n_{0p}^-$ is 
vanishing small and $n_{0p}^+$ is a bell-shaped curve with its maximum 
at $p_3 = 0$. We will proceed as in \cite{Perez} and 
call $p_0 (\gg \sqrt{- 2 M \mu'})$ some characteristic momentum. 
We have then, by assuming $-\mu' 
\ll T$, for the net density  in a small neighborhood of $p_3 = 0$,

\begin{eqnarray}
N_{0 s,v} & = & \frac{e B T}{2 \pi^2 \hbar^2 c}\int_0^{p_0} \frac{d 
p_3}{\sqrt{p_3^2 c^2 + M^2 c^4 \pm e B \hbar c} - \mu} -\int_0^{p_0}
\frac{d 
p_3}{\sqrt{p_3^2 c^2 + M^2 c^4 \pm e B \hbar c} + \mu} \nonumber
\\[1em]
&&\mbox{}\simeq \frac{e B T}{2 \pi^2 \hbar^2 c^2}\int_0^{p_0}
\frac{(M_{\pm}c^2 + \mu) d 
p_3}{p_3^2 c^2 +M_{\pm}^2 c^4 - \mu^2}-  \int_0^{p_0} \frac{(M_{\pm}c^2 -
\mu) d 
p_3}{p_3^2 c^2 +M_{\pm}^2 c^4 - \mu^2}\nonumber
\\[1em]
&&\mbox{}= \frac{e B T}{4 \pi \hbar^2 c}[\sqrt{\frac{M_{\pm}
c^2+\mu}{M_{\pm} c^2-\mu}}-
\sqrt{\frac{M_{\pm}c^2-\mu}{M_{\pm} c^2+\mu}}] \sim \frac{e B T}{4 \pi
\hbar^2 c}\sqrt{\frac{2M_{\pm}}{-\mu'}}  \label{5}
\end{eqnarray}
\noindent
where $N_{s,v} = N_{0 s,v} + \delta N_{s,v}$ and $\delta N_{s,v}$ is the 
density in the interval $[p_0, \infty]$. Actually $\delta N_{s,v}$ is
negligibly small and $N_{s,v}^0 \simeq 
N_{s,v}$. We observe that the contribution of antiparticles can be
neglected as $\mu \to M_{\pm}$.
 Thus  (\ref{5}) leads to
\noindent
\begin{equation}
\mu' = - \frac{e^2 B^2 T^2 M_{\pm}}{8 \pi^2 
N_{0 s,v}^2 \hbar^4 c^2}. ~\label{3}
\end{equation}
We observe 
that $\mu'$ is a decreasing function of $T$ and vanishes for $T = 0$, 
where the "critical" condition $\mu = M_{\pm} c^2$ is reached.
As shown below, in that limit the Bose-Einstein 
distribution degenerates in a Dirac $\delta$ function, which means to have 
all the system in the ground state $p_3 = 0$. To see this, we shall 
rewrite  the momentum density of particles around the 
ground state $p_3 = 0$, $n = 0$ approximately as 
\begin{eqnarray}
n_0 (p_3) &=& \frac{T}{\frac{p_3^2}{2 M_{\pm}} + \frac{e^2 B^2 T^2
M_{\pm}}
{8 \pi^2 \hbar^4 c^2 N_{0 s,v}^2}}\nonumber
\\[1em]
&&\hbox{}=\frac{4 \pi \hbar^2 c N_{0 s,v}}{e B}\cdot \frac{\gamma}{p_3^2 + 
\gamma^2} 
\end{eqnarray}
where 
\[
\gamma = \frac{e B T M_{\pm}}{2 \pi \hbar^2 c N_{0 s,v}} = 
\frac{p_T}{v N_{0 s,v}} = \sqrt{- 2 M_{\pm} \mu'} \]
\noindent
where $p_T = \sqrt{2 \pi M_{\pm} T}$ is the thermal momentum, $v = 
h c\lambda/e B$ the elementary volume cell, $\lambda = h/p_T$ being 
the De Broglie thermal wavelength (observe that $v$ decreases with
increasing $B$). We have 
approximated the Bose-Einstein distribution by one proportional to a 
Cauchy distribution, having its maximum  
$\sim \gamma^{-1}$ for $p_3 = 0$. We have that $\gamma \to 0 $ for $T \to
0 $, but for small fixed $T$,  
$\gamma$ also decreases as $ v N_{0 s, v} $ increases. We remind that in 
the zero 
field case, the condensation condition demands $N \lambda^3 > 2.612$.

\noindent
As a property of the Cauchy distribution one can write approximately,
\begin{equation}
\frac{1}{2} N_{0 s,v} = \frac{ e B}{4 \pi^2 \hbar^2 c} 
\int_{-\gamma}^{\gamma} n_0 (p_3) d p_3. \label{cond}
\end{equation}
Thus, approximately  one half of the total density
is concentrated in the narrow strip of width $2\gamma$ around the 
$p_3 = 0$ momentum. It results that for densities and magnetic fields
large 
enough, if 
we choose an arbitrary small neighborhood of the ground state, 
of momentum width $2 p_{30}$, one can always find a temperature 
$T > 0$ small enough such that $\gamma \ll p_{30}$ and 
(\ref{5}) and (\ref{cond}) are satisfied. The
condensate appears and it is described by the statistical distribution.
Graphs of $n_0 (p_3)$ for some values of $\bar\mu$ are shown in Fig. 2.
%%%%%%%%%
We have also
 \begin{equation}
\lim_{\gamma \to 0} n_0 (p_3)  =  4 \pi^2 \hbar^2 c \frac{N_{s,v}}{e B} 
\delta(p_3). \label{delta}
\end{equation}
Or equivalently,
\begin{equation}
N_{s,v} = \lim_{\gamma \to 0} \frac{ e B}{4 \pi^2 \hbar^2 c} 
\int_{-\infty}^{\infty} n_0 (p_3) d p_3. 
\end{equation}
\noindent
Thus, if $T \to 0$, all the density $N_{s,v}$ falls in the 
condensate, as occur in the zero field case, but here the total density 
is described explicitly by the integral in momentum space. 

One cannot fix any  (small)  value for 
$\gamma$ at which   
the distribution starts to have a manifest  $\delta (p_3)$ behavior; 
and {\it there is 
no} definite value of critical 
temperature for condensation to start to be constrained essentially in the
ground
state. But
the confinement in the $n=0$ Landau state occurs if the ratio
$eB/T$ is large enough to make the average occupation numbers in excited
Landau states negligibly small as compared to the ground state. The
confinement to the $p_3 =0$
momentum state along the field, for a given $B$, is determined by the
ratio  $T/N$. If it is small enough to make 
$\gamma \ll 2 \pi \hbar/L$ 
where $L$ is the characteristic length of
the system along the magnetic field the  condensation takes place
essentially in the ground state.
(these conditions may be achieved in
non-relativistic systems e. g., for masses $\sim 10^{-27}$ g, fields of
order
$10^6$ Gauss, 
$T=10^{-2}$$^{\circ}$K 
and $L \sim 1$cm; for Kaons in a
neutron star, assuming $T=10^{8}$$^{\circ}$K, $N=10^{39}$,
$B=10^{10}$Gauss, $L\sim 10$ Km, it results $\gamma \simeq 10^{-29}$ and
$\gamma \ll 2 \pi \hbar/L \simeq 10^{-33}$. One half of the total density
can be assumed then to be distributed among $10^4$ states; i.e., the
ground
state density is $10^{35}$, which is a macroscopic number ), 
The characteristic order of magnitude of the temperature for having a
macroscopic fraction of the total density in the ground state, if $L$ is
the dimension of the system parallel to the field, is given by $L/\lambda
\simeq vN$, or
\begin{equation}
T_d \simeq \frac{4\pi^2 \hbar^3 c N}{e B M_{\pm} L}.
\end{equation}
 
We observe that for given $N/T$, the quantity $\gamma$ increases with $B$,
that is, the increasing field tends to spread the distribution along
$p_3$, or in other words, to enlarge the density in states close to the
groud state, and if $T \simeq T_d$, condensation in  states neighbor to
the
ground state become also significant.  But for $T \ll T_d$, 
we have almost all the particles in the ground state $n=0$, $p_3= 0$, and
a true condensate exists. 

We return to expression (\ref{5}). It is easy to obtain the infrared 
contribution to the thermodynamic potential as 
\begin{equation}
\Omega = \frac{eBT}{4\pi \hbar^2 c}\sqrt{M_{\pm}^2 - \mu^2}
\end{equation}
and from $\Omega$, we get the energy $U = T\partial \Omega/\partial T
-\Omega$,
 and the specific heat, which as $T \to 0$ tends to the constant value,
\begin{equation}
c_V =\frac{2 e^2B^2 M_{\pm}k^2}{4\pi^2 \hbar^4 c^2 N_{0s,v}},
\end{equation}
\noindent
where $k$ is the Boltzmann constant..

\section{Magnetization and superconductivity}

For the vector field case, the 
magnetization in the condensation limit is positive since all the system 
is in the Landau $n = 0$ state, and
\begin{eqnarray}
{\cal M}& = &- \frac{\partial \Omega}{\partial B}\nonumber
\\[1em]
&&\hbox{} = \frac{e^2 B}{8 \pi^2}\int_\infty^\infty \lim_{\gamma \to 0}
n_0 (p_3) dp_3
\\[1em]
&&\hbox{} = \frac{e N_v \hbar}{2 M_{-} c}. \label{mag}
\end{eqnarray}
\noindent
and we have that the condensate of vector particles leads to
a true ferromagnetic behavior. In particular, if we  write $H_c^{ext} =
0$, we get 
$B = {\cal 
M}(B)$ as the condition for spontaneous magnetization to occur.
If we started at high temperature $T$, where the system have a
diamagnetic contribution coming from the distribution in Landau states,
$n=1,2..$, and a paramagnetic effect due to the spin coupling, we observe
that by lowering $T$ gradually, the system ends in a true ferromagnetic
state. A phase transition without having a critical temperature, as the
"diffuse" ones \cite{Smol}, has taken place.

Usual superconductivity may be understood in principle as a manifestation
of BEC of Cooper pairs, which are spin zero fields. In connection to it
Schafroth \cite{Schafroth} in his study of the analogy between Bose
condensation of scalar charged
particles
and superconductivity,  found a critical
applied field
$B_s = eN_{cs}/2 M_{+}c$, (where $N_{cs} = N_s [1 - (T/T_c)^{3/2}]$, 
$N_s$ is the total density and $T_c$ the critical temperature for normal 
condensation),
such that the
magnetization is
equal and opposite to the applied field. In this case when
the condensate is contained in some body limited by a surface, and the
system 
is placed in an external magnetic field, it reacts 
by creating a surface magnetization
(Meissner effect) which leads to a
magnetic field
equal and opposite to the applied one. The charges inside does not feel
the
external field and remain in the condensate. For field intensities $B >
B_s$ the (zero field) condensate is destroyed. 
Obviously, this is the case
of not very intense fields, in which the system of charges at the
surface are distributed on a large set of Landau
states producing a magnetization which is equal and opposite to the
external 
field, keeping the particles inside in free field states. 
But as the magnetic field grows some
of the population inside starts to accommodate in the ground state, and
for
larger field intensities, the condensate dissapears. However, if the 
magnetic field increases enough, all the system falls in the $n=0$ Landau 
ground state, and if the relation $N/T$ is large enough, the condensation 
becomes manifest again. By calculating ${\cal M}_s= -\partial
\Omega/\partial B$, as the density is (\ref{delta}), we have
\begin{equation}
{\cal M}_s= - \frac{eN_s\hbar}{2 M_{+}c}
\end{equation}

Thus, for growing fields starting from zero, we have usual condensation
(and
superconductivity), up to the Schafroth's critical field. Then appears a
gap where condensation and
superconductivity are not observable, and for field and $N/T$ large enough
condensation (and
superconductivity)  reappears.

 The phenomenon of condensation of scalar and
vector charged particles in strong magnetic fields may have, thus,
especial
interest in connection
with condensed matter physics and cosmology (the early universe) . In
relation with the
first, it has been reported (see i.e \cite{Boebinger}), that the critical
field of some high $T_c$ superconductors
diverges with decreasing $T$, which suggests
the appearance
of re-entering superconductivity in extremely strong magnetic fields.
The vector component of this superconductor, according to our previous
results, behave as ferromagnetic. Thus,  for the 
superconductive phase in extremely strong magnetic fields 
we may expect the appearance of
ferromagnetic-superconductive properties in
those cases in which there is a significant amount of parallel spin
paired  fermions.

\section{Magnetization in the early universe}

We will consider the spontaneous magnetization of vector particles in 
connection with the condensation of $W^{\pm}$ bosons, a phenomenon which
was suggested by Linde \cite{Linde} to occur in superdense matter, and
by one of the present authors and Kalashnikov \cite{Oleg}, to occur near
the electroweak phase transition. The condensation of $W$'s must be 
understood as some chemical equilibrium process, since the $W$'s
interacting with 
the gluonic background would decay in $n$, $p$ or $\pi^{\pm},\pi^0$ pairs.

If we consider at first the large density case, by taking $N_W \simeq 
N_{v c}$, where $N_{v c} = M_W^3 c^3/6 \pi^2 \hbar^3 \simeq 10^{45} {\rm 
cm^{-3}}$, the magnetization would be
of order $B \simeq 10^{19}$ Gauss. One may think as the mechanism 
of arising these huge fields as follows: the microscopic fields inside 
hadrons are of order $10^{15}$ Gauss. Such fields lead the $W$-condensate
to
magnetize, creating a field $10^4$ times larger. Then the 
conditions for spontaneous magnetization in domains arise. Obviously,
${\cal M} $ becomes very large (of order $B_c$) for fields near the
critical value 
$B_c \simeq M^2_W c^4/e \hbar c 
\simeq 
10^{24}$ gauss, since for $B \geq B_c$  the system becomes unstable
leading 
to imaginary energy eigenvalues.
\cite{Nielsen}. Near  the symmetry restoration temperature as $M =
M(T) < M(0)$  the critical field occurs for smaller values of $B$.

The case of temperatures near the symmetry restoration lead to very 
interesting results. As shown in \cite{Oleg},\cite{Chai}, condensation of 
transverse vector $W$ bosons occurs for arbitrary small densities, and
taking into 
account that the strong local magnetic fields would magnetize the 
condensate, we may think that (\ref{mag}) becomes singular, (for
$M_v (T)= \sqrt{M^2_W (T) - eB\hbar/c^3} \to 0$, leading to 
infinite 
large magnetization of the medium, due to the vanishing of the 
(transverse, since the longitudinal modes acquire a Debye mass $\sim g T$) 
vector boson mass. One may think that this would lead to
instabilities also, but we must remember that the zero mass of the $W_s$
for $T > T_c$ is only an approximate result first because actually the
mass is
expected
to be of 
order $g^2 T$ \cite{Linde1}, second, because the self-magnetization
$B={\cal M}(B)$ that would arise, leads to a mechanism preventing
the divergence to occur. In any case, the existence of a condensate
in presence of  strong hadron (or quark) fields may lead to the arising of 
a ferromagnetic behavior leading to extremely 
large fields, of order $B \leq B_c$,
at
temperatures $T \sim T_c$. The existence of such large fields of order 
$B_c$ , as
fluctuating fields on a scale $M_W^{-1}$, was first suggested by
Vachaspati \cite{Vachaspati}. The same order has been estimated in
\cite{Baym} starting from the idea of some sort of
equipartition of
magnetic and radiation energies that may occur in a magnetically turbulent
environment in the bubble formation arising from an electroweak first
order phase transition.

\section{Conclusions}
We conclude that Bose-Einstein condensation of charged particles in a
strong magnetic field is possible and lead to several new and interesting
phenomena, as it is the occurrence of  phase transition in presence of an 
external magnetic field, not
having a critical temperature. For low field intensity we have usual
condensation,
and for very strong fields, condensation is manifest again. The
condensate in the strong magnetic field suggests the existence of
superconductivity in extremely strong magnetic
fields and the existence of a ferromagnetic-superconductive phase. This
has
interest in condensed matter physics. In astrophysics and cosmology we
have also interesting consequences. It gives support to the conjectured
existence of
superfluid and superconductive phases in neutron stars \cite{Thor}. It
suggests also that
at the electroweak phase transition, extremely strong magnetic fields
may arise as a consequence of condensation and self-magnetization effects 
of
the medium.

The authors would like to thank J. G. Hirsch for very important and
detailed discussions and suggestions. They thank also R. Baquero and G.
Gonzalez for fruitful
comments. One
of the authors (H.P.R.) thanks H. Rubinstein and G. Andrei Mezincescu for
important discussions
and M. Chaichian for hospitality in the High
Energy Division of the Department of Physics, and acknowledges
the financial support of the Academy of Finland under the Project No.
163394.

\newpage
\begin{center}
Captions
\end{center}

Figure1. {\sl The 
density in momentum space of the Bose gas for $T > T_c$ has a minimum at
$p_3 = 0$ and a maximum at some $p_3 \neq 0$. As 
temperature is 
decreased, the maximum approaches to the value $p_3 = 0$ and at $T = T_c$ 
it is rigorously on it, the minimum disappears. The curves a, b, c, d
correspond respectively to $\bar\mu$ = 1, 0.1, 0.001 and 0.}

Fig 2. {\sl In the magnetic field case, for $e B \hbar/mc T \gg 1$,
the system is confined to the $n = 0$ Landau quantum state. The maximum 
of the density in the momentum component along the magnetic field is 
located at zero momentum at any temperature, and for very low $T$, it has 
a peaked $\delta$-like form. Here curves d, e, f correspond to $\bar\mu$ = 
1, 0.1 and 0.001.}


\begin{thebibliography}{99} 
\frenchspacing
\bibitem{Chai} M. Chaichian, R. Gonzalez Felipe, H. 
Perez Rojas, Phys. Lett. {\bf B 342} (1995), 245.
\bibitem{Schafroth} M. R. Schafroth, Phys. Rev. {\bf 100} (1955) 
463. 
\bibitem{May} R. M. May, J. Math. Phys. {\bf 6} (1965) 1462.
\bibitem{Daicic} J. Daicic, N. E. Frankel, and V. Kowalenko, Phys. 
Rev. Lett. {\bf 71} (1993) 1779.
\bibitem{Gailis}J. Daicic, N. E. Frankel, R. M. Gailis and V. Kowalenko, 
Phys. Rep.{\bf 237} (1994) 63..
\bibitem{Toms} D. J.Toms, Phys. Rev. Lett. {\bf 69} 
(1992), 1152; Phys. Rev. {\bf D 47} (1993), 2483;
Phys. Lett. {\bf B 343} (1995), 259
\bibitem{Elm} P. Elmfors, P. Liljenberg, David Persson, Bo-Sture
Skagerstam,
Phys. Lett.{\bf B 348} (1995), 462.
\bibitem{Toms1}  K. Kirsten and D. J. Toms, {\it Phys. Rev. } {\bf A 54}
(1996), 4188.
\bibitem{Kirsten} K. Kirsten and D. J. Toms,  Phys. Lett. {\bf B} 368
(1996) 119.
\bibitem{Pathria} R. K. Pathria, {\it Statistical Mechanics}, Pergamon 
Press, Oxford, (1978).
\bibitem{Kagan} Yu. Kagan, N.V. Prokof'ev and D.V. Svistunov, {\rm
cond-mat/9908378}
\bibitem{Landau} L. D. Landau and E. M. Lifshitz, {\it Statistical
Physics}, 
3rd. Ed., Part 1, Pergamon Press (1986), New York.
\bibitem{HPerez} H. Perez Rojas, Phys. Lett.{\bf B  379} (1996) 148.
\bibitem{Hugo} H. Perez Rojas,  Phys. Lett. {\bf A 234} (1997), 13.
\bibitem{Fradkin} E.S. Fradkin, {\it Proc. of the P.N. Lebedev Physical
Inst. No.}{\bf 29}, Cons. Bureau (1967). 
\bibitem{Perez} H. Perez Rojas, Acta Phys. Pol. {\bf B 17} (1986), 861.
\bibitem{Smol} G. A. Smolenski and V. A. Isupov, Sov. Journal of 
Techn. Phys.{\bf 24} (1954) 1375; R. L. Moreira and R. P. S. M. Lobo,
 Jour. Phys. Soc. Japan {\bf 61} (1992), 1992.
\bibitem{Boebinger} M. S. Osovski, R.J. Soulen Jr., S.A. Wolf, J.M. Broto,
H. Rakoto, J.C. Oussel, G. Coffe, S. Askenazi, P. Pari, I. Bozovic, J.N.
Eckstein and G. F. Virshup, {\it Phys. rev. Lett.} {\bf 71} (1993) 2315; 
Yoichi Ando , G.S. Boebinger and A. Passner, {\it
Phys. Rev. Lett} {\bf 75} (1995) 4662, M. Rasolt and Z. Tesanovic, Rev.
Mod. Phys. {\bf 64} (1992), 709.
\bibitem{Linde} A. D. Linde,  Phys. Lett. {\bf 86 B} (1979), 39.
\bibitem{Nielsen}  N.K. Nielsen and P. Olesen, {\it Nucl. Phys.} {\bf B
144}
(1978) 376; J. Ambjorn, N.K. Nielsen and P. Olesen, {\it Nucl. Phys.} {\bf
B
310} (1988), 625.
\bibitem{Oleg} O. K. Kalashnikov, H. Perez Rojas,  Kratkie Soob. po 
Fizike, Lebedev Inst. Reports, Allerton Press {\bf 2} (1986) 2; H. Perez 
Rojas, O.K. Kalashnikov, Nucl. Phys.{\it B 293} (1987), 241.
\bibitem{Linde1} A. D. Linde,  Rep. Prog. Phys. {\bf 42} (1979), 389.
\bibitem{Vachaspati} T. Vachaspati, {\it Phys. Lett.} {\bf B 265} (1991), 
258.
\bibitem{Baym} G. Baym, G. Bodecker and D. Mc Lerran, Phys. Rev. {\bf
D 53}
(1996), 662.
\bibitem{Thor} V. Thorsson, M. Prakash and J.M. Latimer,  Nucl.
Phys. {\bf A 572} (1994), 693.
\end{thebibliography}
\end{document}